\documentclass[mypaper]{myaa}
\usepackage{graphicx}
\usepackage{natbib}
\usepackage{balance}
\bibpunct{(}{)}{;}{a}{}{,} 

\idline{0}{1}

\begin{document}

\title{Beryllium  abundance in turn-off stars of NGC~6752
\thanks{Based on observations collected at the ESO VLT, Paranal  
Observatory, Chile , program 075.D-0807(A) }}
\subtitle{}
\author{L. Pasquini\inst{1}, P. Bonifacio\inst{2,3,4}, S. Randich\inst 
{5}, D. Galli\inst{5}, R. G. Gratton\inst{6}, B. Wolff \inst{1}}

\offprints{lpasquin@eso.org}

\institute{European Southern Observatory, Garching bei M\"unchen,  
Germany
\and
CIFIST Marie Curie Excellence Team
\and
Observatoire de Paris, GEPI, F-92195 Meudon Cedex, France
\and 
INAF--Osservatorio Astronomico di Trieste, Trieste, Italy 
\and
INAF--Osservatorio Astrofisico di Arcetri, Firenze, Italy
\and
INAF--Osservatorio Astronomico di Padova, Padova, Italy }

\titlerunning{Be in NGC~6752}
\date{Submitted: Accepted}
\abstract
{}
{To measure the beryllium abundance in two TO stars of the Globular Cluster
NGC 6752, one oxygen rich and sodium poor,  the other  presumably   
oxygen poor and sodium rich. Be abundances in these
stars are used to put on firmer grounds 
the hypothesis of Be as cosmochronometer and to investigate the formation 
of Globular Clusters.  }
{We present near UV spectra with resolution R$\sim 45000$ obtained
with the UVES spectrograph on the 8.2m VLT Kueyen telescope, 
analysed with spectrum synthesis based on
plane parallel LTE model atmospheres.}
{Be is detected in the O rich star with
log(Be/H)=-12.04 $\pm$0.15, while  
Be is not detected in the other star for which we obtain the
upper limit log(Be/H)$<$-12.2.
A large difference in nitrogen  abundance (1.6 dex)
is found between the two stars.
}
{The Be measurement is compatible
with what found in  field stars with the same [Fe/H] and [O/H].
The 'Be age' of the cluster is found to be 13.3 Gyrs,
in excellent agreement with the
results from main sequence fitting and stellar evolution. 
The presence of Be confirms the results previously obtained for
the cluster NGC~6397 and supports the 
hypothesis that Be can be used as a clock for the early formation of the Galaxy. 
Since only an upper limit is found for the   star with low oxygen abundance,
we cannot decide between competing scenarios of Globular Cluster formation, 
but we can exclude that 'polluted' stars are substantially
younger than 'unpolluted' ones.
We stress that the Be test might be the only measurement capable of distinguishing 
between these scenarios. 
}
\authorrunning{L. Pasquini et al.}
\titlerunning{Be in NGC~6752}
\maketitle


\section{Introduction}

In a recent paper we have successfully tested the use of beryllium  
(Be) as
a cosmochronometer for the early phases of Galactic evolution
(Pasquini et al. 2004). Be is produced by the interaction
of Galactic cosmic rays (GCRs) with the interstellar medium,
through the spallation of heavy element nuclei, most noticeably: carbon, nitrogen, and oxygen.
Due to its origin, and in particular to the fact that GCRs are  
generated
and transported globally on a Galactic scale,
Be is expected to
be characterized by a smaller dispersion than the products of type II
supernovae, whose abundances in the early Galaxy are affected by
the dispersed character of star formation and inefficient mixing  
of gas.
This led to the suggestion that Be represents a ``cosmic clock"
(Suzuki et al. 2001; Beers et al. 2000). To test this hypothesis
Pasquini et al. (2004) carried out the first measurements of Be
abundances in a globular cluster (GC); namely they observed Be in
two turn-off (TO) stars of the metal poor ([Fe/H]$\sim -2$) GC  
NGC~6397,
for which an
independent age estimate was available.
Be was detected in both stars at a level consistent with that
of stars in the
field with the same [Fe/H] abundance.
By comparing their Be values
with models of Galactic 
evolution of Be as a function of time, they concluded
that the cluster formed about 0.2-0.3~Gyr after the onset of star
formation in the halo, in very good agreement with the cluster age
derived from main sequence fitting. In order to prove that this  
result
is not fortuitous, it is extremely important to carry out the same
test on additional GCs.

As discussed by Pasquini et al. (2004), Be and Li measurements
in GCs provide as well important constraints
on the hotly debated issue of cluster formation.
Detailed studies of chemical abundances in
GC stars have revealed that
anomalies are present in all the clusters studied, showing that
they are not a homogeneous populations as far as the
chemical composition of the stellar atmospheres is concerned
(e.g. Gratton et al. 2004 and references therein).
More specifically, whereas GCs appear to be extremely
homogeneous in Fe and Fe-peak elements,
star-to-star variations are seen for those elements
involved in H-burning at high temperature.
This behaviour is not present among field stars, which do not exhibit
significant
star-to-star variations in abundance ratios like [O/Na] or [Mg/Al]
at a given overall metallicity.
This intriguing abundance pattern, and in particular the fact that
abundance anomalies are seen not only among evolved cluster stars
but also in unevolved stars at the TO, suggests that
part of the gas forming the stars that we observe today has been 
processed and ejected by a 
previous generation of stars. When considering that GCs are almost  
as old as
the age of the Universe, this implies that this hypothetical previous
generation might
have been part of the elusive metal-free Population III,
the first stars formed.
By studying the current chemical composition of GC stars we may
therefore try to
determine the composition of the ejecta of the first stars and
to understand their basic characteristics (mass, chemical  
composition).

In a series of papers we have investigated the light elements
Li and Be in TO
stars of GCs (Bonifacio et al. 2002; Pasquini et al. 2004, 2005;  
Bonifacio
et al. 2007, in preparation), showing that these elements are indeed
very useful to investigate the process of GC formation.
We recall that $^7$Li and $^9$Be are both destroyed at relatively low
temperatures ($\sim$2.5 and $\sim$3.5 million K respectively) in stellar  
interiors.
These temperatures are significantly
below those at which typical reactions responsible for the
chemical anomalies in GCs occur. Thus, in the hypothesis that
a fraction of GC stars is born from the ejecta of a previous
generation of stars, these stars should in principle show low Be and
Li abundances.
Note, however, that Li and Be form through different mechanisms.
Whereas, as mentioned,
$^9$Be can only be produced by GCR spallation on the ISM,
$^7$Li is produced during Big Bang Nucleosynthesis,
and the Galaxy is later enriched by
several mechanisms, such as GCRs spallation, AGB stars, and,  
possibly,
Novae (e.g., Travaglio et al. 2001).

Our observational results can be summarized as follows: the
Li abundance is constant in NGC~6397 (Bonifacio et al. 2002),
while Li is found to vary from star-to-star in NGC~6752 and 47 Tuc.
In both these GCs,
Li abundances are correlated with O and anti-correlated with Na  
abundances
(Pasquini et al. 2005; Bonifacio et al. 2006).

As for Be, the mere existence
of Be in in the two TO stars of NGC~6397 suggests that the gas
which formed the stars we now
observe must have been sitting for at least a
few hundred million years in the ISM exposed to
the GCR spallation before the 
stars formed.

NGC~6397 could, on the other hand, be a somewhat 'special' case,  
because
it is one of the
GCs where chemical anomalies are present at the lowest level, and,  
as mentioned
above, the Li abundance is strikingly constant.
It may therefore be questioned whether  this
cluster is a fair representative of all GCs. NGC~6752, on the  
opposite,
is a prototype of
GC with large chemical anomalies; these anomalies have been
found all over the colour-magnitude diagram;
the Na-O anticorrelation has been discovered among main sequence
stars (Gratton et al. 2001,
hereinafter G01), as well as Li - Na anticorrelation (Pasquini et  
al. 2005).
This cluster is
therefore a very good prototype for anomalies; it has an
intermediate  metallicity ([Fe/H]=$-1.48$,
G01), and its age is virtually the same of
NGC~6397 according to main sequence fitting
(Gratton et al. 2003).

\section{Sample Stars and Observations}

Stars at the TO of NGC~6752 have been studied extensively for  
chemical abundances
(Gratton et al. 2001, 2003, Carretta et al. 2005, James et al.  
2004, Pasquini et al. 2005).
We selected two stars from the G01 sample having different  
composition; star 4428 is
representative of the O-rich, Li rich component, while the star  
200613 is representative of the
  Li poor component and, by inference, should be also 
oxygen poor . By observing stars at the extremes of the  
chemical distribution, we
aimed at maximizing the chances of observing possible differences  
in Be abundance.

We  recall that the Be lines are located at 313 nm and the Be  
observations of GC
turnoff stars are very challenging, because the spectral region is  
crowded, the Be lines tiny
and the UV flux of these relatively cool stars is low. When adding  
the relatively high
absorption introduced by the earth's atmosphere, it  clearly  
emerges that the TO stars
of even the closest GCs are very faint
 for
observations of Be lines. NGC~6752 TO  
stars are almost one
magnitude fainter in the V band than the NGC~6397 objects, but the  
relatively low reddening
of the cluster (E(B-V) = 0.05: Gratton et al. 2003) allows a  
substantial flux recovery at
the UV wavelengths, so that these observations, even if very  
demanding, are feasible with
UVES at the  VLT .

Table 1 summarizes the characteristics of the stars, including the  
abundances of the single
elements, as derived from the literature. The stellar parameters  
listed in Table 1 are those
adopted in our spectroscopic analysis. Only abundances of those  
elements which are known to
vary from star to star (O, Na, Li) are listed in Table 1; we also  
add in the table the results
for Be and N as resulting from this paper. The reader can find  
additional element abundances
in G01, James et al. (2004), and Carretta et al. (2005). 

\begin{table*}
\caption{NGC~6752 sample stars, their atmospheric parameters and
abundances. The atmospheric parameters,  [Fe/H], [O/H] and [Na/Fe] are
from G01;  Log(Li/H) from Pasquini et al. (2005), 
[N/Fe] and $\log({\rm Be/H})$ from this work. }

\begin{tabular}{lllllllll}
\hline
Star  & T$_{\rm eff}$ & $\log g$   & [Fe/H] & [O/H]   & [Na/Fe]   
&  $\log({\rm Li/H})$ & [N/Fe] & $\log({\rm Be/H})$  \\
         &           &           &           &     &            
&                   &       &         \\
4428     & 6226      &  4.28     &  $-1.52$ & $-1.28$   &  $-0.35 
$  & 2.50             &  0    & $-12.04 \pm 0.15$   \\
200613   & 6226      &  4.28     &  $-1.56$ & $ / $   &  $-0.09 
$  & 2.13             &  1.6  & $< -12.2 $   \\
\hline
\end{tabular}

\end{table*}

The observations were carried out in service mode at the VLT
observatory in several runs during summer 2005 with the UVES
spectrograph (Dekker et al.~2000). We used a $1^{\prime\prime}$ slit
providing a resolving power of 45,000. The blue CCD was binned
to minimize the CCD read out noise.
In addition, all observations were taken at low airmass, to limit
the atmospheric absorption, which is particularly high at these
UV wavelengths. A total of 10 exposures of 90 minutes each were
obtained per star. The observations were reduced using recipes from the  UVES
pipeline (Ballester et al.~2000).
Given the fairly low S/N ratio of each observation in the Be region,
we have been especially careful in reducing the data. The two major
difficulties were identified in the correct estimate of the  
background and in the
elimination of the 'cosmic ray' events. We have therefore used two  
different
reductions, either using the standard calibration and pipeline, or  
using the
D lamp for flat fielding, which gives better spectra in the blue.  
This lamp, however,
has some emission lines in its spectrum;
it can be used in the spectral
region of the Be lines, but not over the whole spectrum.

The reduced spectra were averaged
leading to a S/N ratio in the Be region of 20 and 10 per 0.025  
{\AA} pixel for stars
4428 and 200613 respectively. The spectra in the Be region, together
with our synthetic best fits (cfr. next section) are shown in  
Figure 1.

\section{Abundance analysis}

The method used has been described in Pasquini et al. (2004).
A full synthesis of the region around the Be doublet has been  
performed
using the line list given in Pasquini et al.(2004).
For each star the best abundance was found by a $\chi^2$ fit to  
the whole
feature. The best fits, shown in Fig.~1, correspond to Be  
abundances of
$\log(Be/H)=-12.04$ for the O-rich, Li-rich star 4428, while only  
an upper limit of $\log(Be/H)<-12.2 $ could be
found for star 200613.
Fitting errors were estimated with Monte Carlo simulations  
assuming a Poisson
noise.  The dispersion around the mean in 1000 Monte Carlo samples  
was
of 0.04 dex for star 4428. These should be viewed as
lower limit to  the errors, 
since at these low S/N ratios other  sources
of non-Poisson noise (e.g.  shot-noise) could be important. To  
these errors
associated to the noise in the data one must add the errors which
derive from the uncertainty in the analysis procedure (atmospheric  
parameters,
etc.). Since we
are using lines of singly ionized Be, surface gravity is the  
parameter
which most affects the Be abundance computation.  The values of $ 
\log {g}$
have been estimated by comparing the position of the stars in the
color-magnitude diagram with theoretical isochrones.
The uncertainty in such determination
is dominated by possible errors in the distance moduli; we can  
then safely assume an
error of 0.15~dex in $\log {g}$ that translates into an
uncertainty of 0.07~dex in the Be abundance.
The error arising from a change of 100~K in the effective  
temperature is about two times lower: 0.03 dex.
If we add  these values to the errors due to noise
(0.04~dex and 0.11~dex, respectively), we obtain an estimate of the
total error on the Be abundance of 0.15~dex.
Note that this error does not include possible systematic errors
due to shortcomings and inadequacies of our modeling (model
atmospheres, atomic data, etc.).

We recall that, while G01 used the same temperature for all NGC~6752 stars, 
given the strong dependence of Li on T$_{eff}$, 
 Pasquini et al. (2005)  investigated the effects of adopting an 
 additional temperature scale, and derived  for these stars  
 T$_{eff}$ from the $b-y$ color as well, finding  6013 K for star 4428 and 
5948 K for star 200613.
These temperatures, when entered
into the ischrones of Straniero et al. (1997) 
imply slightly higher gravities for both
stars, namely  $\log g = 4.43$ for star  4428
and 4.46 for star  200613.The combined effect of this change in atmospheric
parameters is an increase of the Be abundance by 0.12 dex with respect to the values given in Table 1.  

\begin{figure*}
\begin{center}
\resizebox{\hsize}{!}{\includegraphics[clip=true]{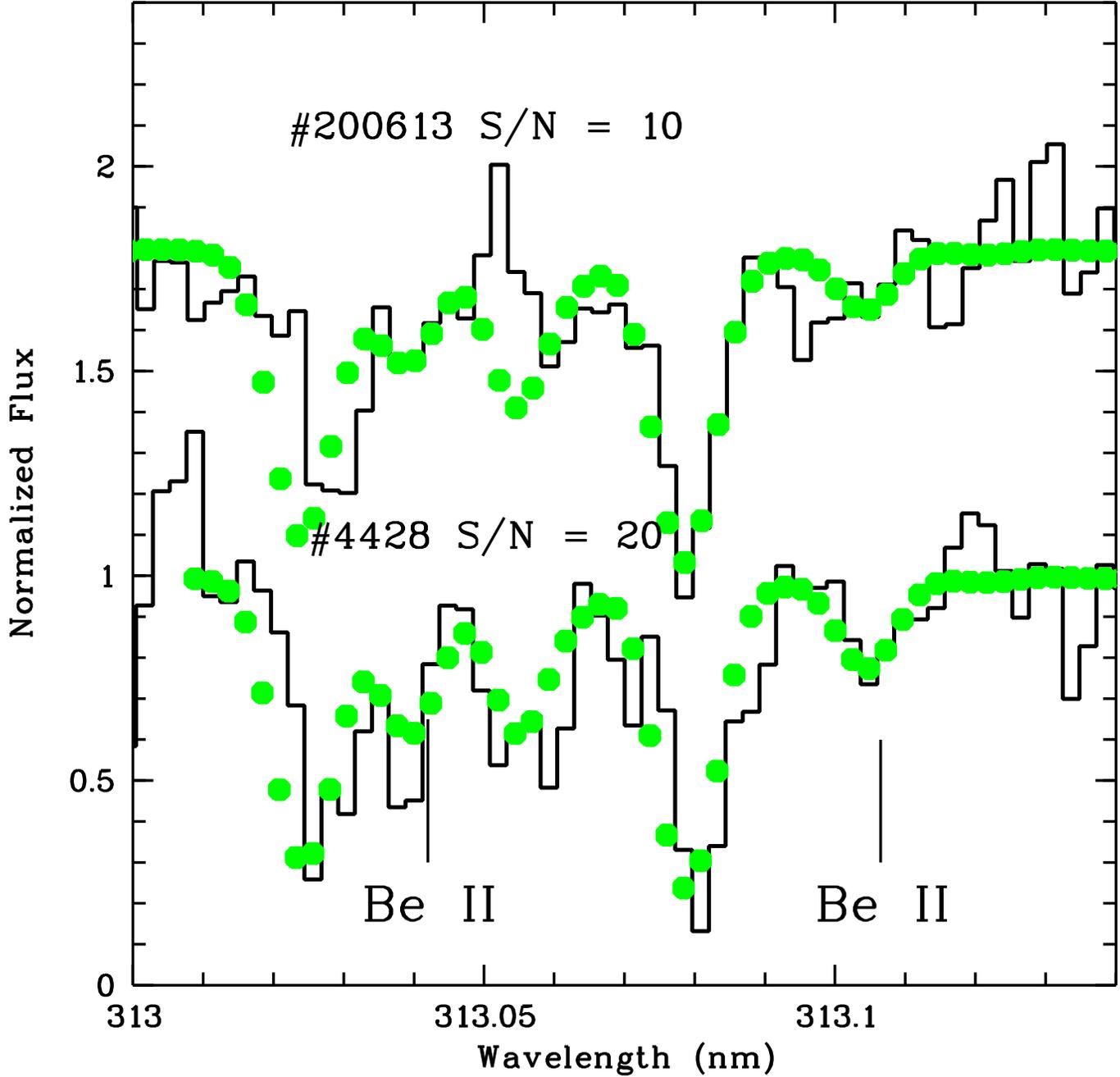}}
\end{center}
\caption{UVES spectra of the two turn-off stars 200613 and 4428 of  
the GC NGC~6752. The dotted lines correspond to the 
Be abundances $\log$(Be/H)$=-12.2$ (upper limit) for 
star 200613 and $ \log$(Be/H)$=-12.04$
for star  4428.}
\end{figure*}

We have also been able to measure the N abundance in these two
stars, confirming that a very high N  difference is present among them.
This can be easily seen in Figure 2, which
shows the NH band at 336 nm for the two stars. Clearly their N  
abundance is different.

We estimate the nitrogen abundance by spectral synthesis of two NH UV
bands, around 336 nm and around 340.5 nm.  Given the complexity of  
the bands it is not possible to carry out a formal fitting procedure  
of the whole spectral region. We have therefore inferred a ``best  
abundance''
by fitting the most prominent features.  Despite several  
systematic and
random uncertainties, we can compute a difference of about
1.6 dex in N abundance among the two stars. For star 200613 the
value of [N/Fe]=1.6 is compatible with what found by Carretta et  
al. (2005) ([N/Fe]=1.7), who used different spectra and 
different spectral features (CN bands).
As far as star 4428 is concerned, the value found ([N/Fe]=0) is
1 dex below that found by Carretta et al. (2005), which should be  
considered, however, as an upper limit rather than a detection ([N/Fe]=1.1).

The discrepancy between the  two [N/Fe] abundance of the  
'unpolluted' star 4428 is critical, since the
evolution of nitrogen in GCs is one of the key data  
to understand the formation process of  the clusters.
In fact, while a value of [N/Fe]=0 is  perfectly compatible with  
that found in metal-poor field stars, 
a value of [N/Fe]=1 (as determined by Carretta et al.) 
 would be definitely  
much higher than that observed in un-evolved stars
in the field,  and it would point out to the need for a different  
source of N production in the
formation process of GC. 
However, a high N abundance is clearly  incompatible with the present
spectra of this star.

Since the carbon isotopic ratio is a very sensitive
diagnostic of mixing in stellar material which
correlates well with [N/Fe] ratios (Spite et al. 2006),
it would therefore be interesting to cross-check
the inference derived from N abundances with 
carbon isotopic ratios.
Unfortunately we do not have $^{12}$C/$^{13}$C ratios
for these two stars, nor for any of the 
dwarf stars in this cluster since all the 
$^{13}$CH features are too weak 
to be measured (Carretta et al. 2005).

An unusually high N  value was found in one NGC~6397 unpolluted  
star (Pasquini et al. 2004)
and a general N overabundance in GC has been postulated, for  
instance, by Li and Burstein
(2003). [N/Fe] among field stars has been studied in the last  
years by several groups,
which found that [N/Fe] is approximately constant with metallicity  
in the range $-3.0 \le
$[Fe/H]$\le 0.4$, as first shown by Carbon et al. (1987) and recently
confirmed by Israelian et al.~(2004) and Ecuvillon et al.~(2004).
According to our results, in NGC~6752, at odds with NGC~6397, the  
'unpolluted' star 4428 does not show
evidence for strong N overabundance with respect to field stars; 
also for this element, abundances in field and 'unpolluted' stars of NGC~6752 
are extremely similar. 

\begin{figure}[]
\begin{center}
\resizebox{\hsize}{!}{\includegraphics[clip=true]{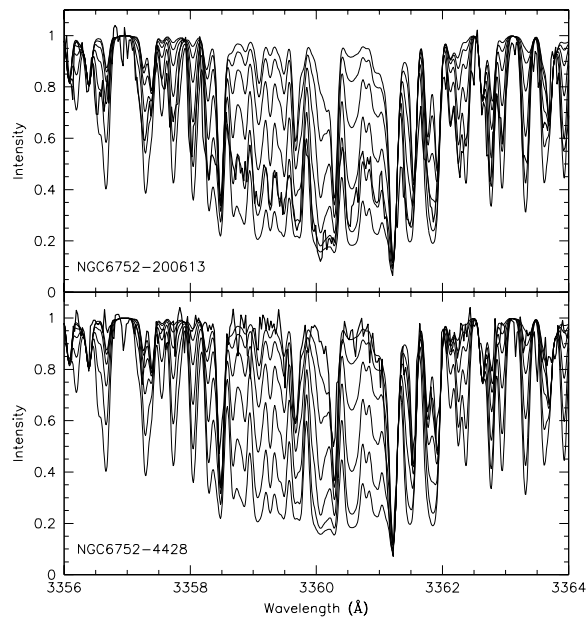}}
\end{center}
\caption{UVES spectra of stars 200613 and 4428 around  the NH band at
336~nm. The band is much stronger in star 200613,
indicating a strong N overabundance. Solid lines are the observed
spectra; thin lines are synthetic spectra computed with parameters  
appropriate
to the program stars, and N abundances of [N/Fe]=0, 0.4, 0.8, 1.2, 1.6  
and 2.0.}
\end{figure}

\section{Discussion}

\subsection{ The age of NGC~6752 and comparison of Be in GC with  
field stars}

As mentioned in Sect.~1, the first detection of
Be in a GC
offered the opportunity to test the theoretical proposal that Be  
could
be used as a cosmochronometer for the early evolution of the Galaxy.
Adopting a standard model of GCRs  spallation for the production
of Be in the Galaxy, Pasquini et al. (2004) found an excellent  
agreement
between the ``Beryllium age'' of NGC~6397 (13.4--13.5~Gyr, assuming a
Galactic age of 13.7~Gyr), and the age of the cluster derived by
Gratton et al.~(2003) from main sequence fitting and theoretical
isochrones ($13.5 \pm 1.1$~Gyr).  The use of Be as a cosmochronometer
has been further exploited by Pasquini et al. (2005) to study the
halo-thick disk formation, finding some evidence for a faster star
formation in the halo and a slower one in the disk.
The detection of Be in the stars of a second GC, NGC~6752,
characterized by a higher metallicity than NGC~6397, and presenting
strong chemical inhomogeneities, is expected to provide additional
constraints to the evolution of Be in the early Galaxy.

\subsection{ The age of NGC~6752 and comparison of Be in GC with  
field stars}

It is clear that the use of Be as cosmochronometer requires a wider
observational support.  In this sense observations of additional
GCs and new comparison between ``Be'' and ``evolutionary'' ages
are necessary steps.

To this goal, we will adopt as the representative Be abundance of
NGC~6752 the value of $\log({\rm Be/H})=-12.04\pm 0.15$ measured in  
star 4428,
which is the prototype of ``unpolluted'' stars in this cluster.  With
its high Li and O, and low Na, the atmosphere of this star is in fact
expected to be composed mainly - if not entirely - of ``unprocessed''
gas (cfr. next section).

In Figure 3 we show the Be abundance vs. time according to the same
Galactic chemical evolution model adopted in Pasquini et al.~(2005)
(from Valle et al. 2002), together with the measured values of Be in
NGC~6397 and NGC~6752.  The assumed age for both clusters is the
``isochrone age'' derived by Gratton et al.~(2003), $13.5\pm 1.1$~Gyr
and $13.4\pm 1.1$~Gyr for NGC~6397 and NGC~6752, respectively.   
The age
difference of 0.1~Gyr between the two clusters 
is much smaller than the error on the age determination. As
in the case of NGC~6397, also for NGC~6752 the agreement with the  
model
predictions is excellent, reinforcing the validity of Be as an
alternative (or additional) age estimator.  A direct comparison  
between
the observed and predicted abundance of Be in NGC~6752 would imply a
``Be age'' of 13.3--13.4~Gyr for this cluster, or about 0.1~Gyr less
than the age of NGC~6397.

As we considered somewhat fortuitous the extremely good agreement
between the ``Be'' and ``isochrone''ages for NGC~6397, we also  
consider
fortuitous the excellent agreement between the two age determinations
for NGC~6752.  It is worth noticing however, that, although the
determination of the absolute value of the age of each cluster may be
affected by large uncertainties with either method, the agreement in
the relative age difference can be considered as a more robust
result, as the effects of systematic errors should cancel.

\begin{figure}
\begin{center}
\resizebox{\hsize}{!}{\includegraphics[angle=0]{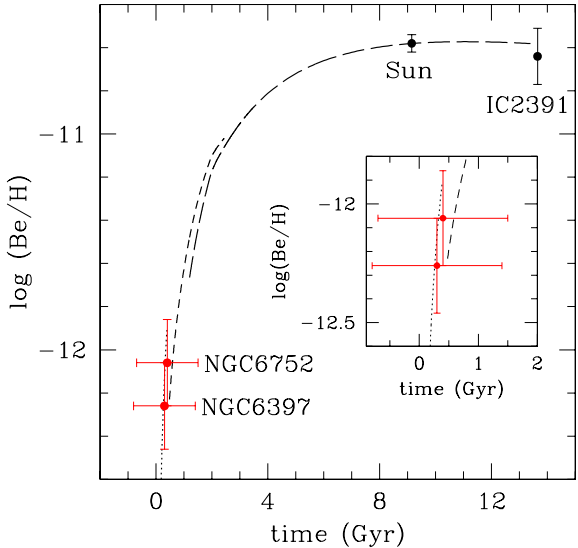}}
\end{center}
\caption{
Evolution of Be with time in the Galaxy according to a
three-zone Galactic chemical evolution model (Valle et al. 2002). The
three curves refer to the halo, thick disk, and thin disk. The data
points show the Be abundance in the young open cluster IC2391 (age
$55$~Myr), the Sun (age $4.5$~Gyr), the globular clusters NGC~6397
(isochrone age $13.5\pm 1.1$~Gyr) and NGC~6752 (isochrone age
$13.4\pm 1.1$~Gyr). The model result is normalized to the solar  
meteoritic
abundance. The inset shows a detail of the early Galactic evolution.
The Be abundance of IC2391 is from Randich et al. (2002);  
the Be abundances
of the two GCs are from Pasquini et al.~(2004)
for NGC~6397 and from this work for NGC~6752. The isochrone ages
are from Gratton et al.~(2003).}
\end{figure}

Pasquini et al. (2004) noticed that the Be abundance of the  
NGC~6397 stars is
fully consistent with the well-known Be vs. Fe trend (Gilmore et  
al. 1992;
Molaro et al. 1997; Boesgaard et al. 1999 -hereafter B99).
In Figure 4 we show the Be vs. Fe relation for the
field stars from B99, together with the NGC~6397 points and the  
NGC~6752 value
determined in this work. Figure 5 is similar, but Be is plotted as
a function of O
rather than Fe. Note that, as discussed by
Pasquini et al. (2004), O
abundances of field (B99) and cluster (G01) stars are on the same  
scale.

The agreement between the cluster and the field stars is impressive:
they are indeed indistinguishable in the Be vs. Fe
plane. In the Be vs. O plane the situation is slightly different,  
because the GC formation process depletes O; this point
will be discussed in the next session. However, insofar  as
'unpolluted' cluster stars
are considered, they are basically indistinguishable from field  
stars not only as far as
nucleosynthesis elements are concerned, but this similarity extends 
to primordial nucleosynthesis products (Li) and to 
elements produced by GCRs spallation (Be).  

A very simple conclusion is that Be production (or more precisely,  
the Be/Fe ratio)
seems to ignore the environment where the stars are born.
We believe that this is a strong evidence in favour to the fact  
that the Be production in the early Galaxy is largely independent of the local  
ISM abundances; rather it reflects some global process, as we proposed in our  
previous work.

The  approximately linear and tight relation between Be an Fe  also  
suggests that their
production sites are linked, and that they are also
fairly independent of the formation site in the early Galaxy.
The simplest explanation is that they are produced by the same
type of events: supernovae, and that the medium was quite
homogeneous at a given time, as far as Fe and Be are concerned.

It will be very difficult to prove the unicity of this solution,  
but the data collected so far are compatible with it. One test could be the  
observations of the Be lines for many stars of different populations
and similar [Fe/H]; would the trends found by Pasquini et al. (2005) 
be confirmed by larger samples of stars, not only the hypothesis of Be as 
a clock  would be re-inforced, but also new insights about the mechanisms of halo 
and disk formation would be gained.

\begin{figure}[ht]
\begin{center}
\resizebox{\hsize}{!}{\includegraphics[angle=270,clip=true]{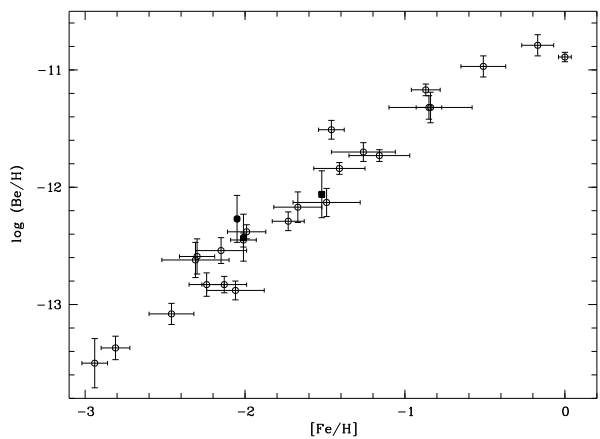}}
\end{center}
\caption{Be abundance vs. Fe 
for the NGC~6397 ({\it filled circles}) and the NGC~6752 TO star ({\it filled square})  
and the star from B99,
({\it  open circles}). The stars in the clusters are indistinguishable from the field   stars. }
\end{figure}

\begin{figure}[ht]
\begin{center}
\resizebox{\hsize}{!}{\includegraphics[angle=270,clip=true]{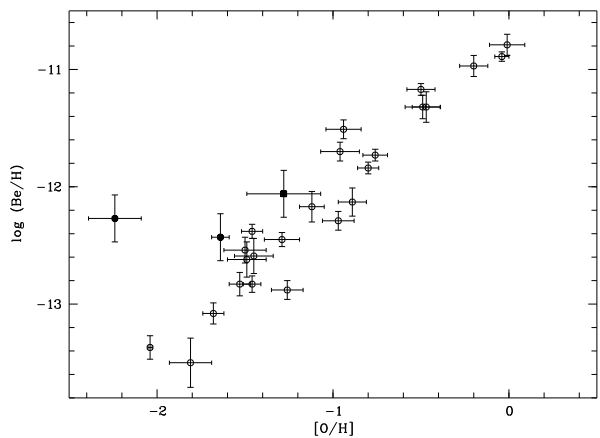}}
\end{center}
\caption{Be abundance vs. O
for the NGC~6397 ({\it filled circles}) and the NGC~6752 TO stars ({\it filled square})  
and the star from B99,
({\it  open circles}). The 'unpolluted',
O-rich stars in the clusters closely follow the field   stars. }
\end{figure}

\subsection{Beryllium and globular cluster formation}

Our Be and N observations must be inserted in the general context
of cluster chemical anomalies (see e.g. Gratton et al. 2004). In  
particular (see e.g. the discussion in Bonifacio et al. 2002, Pasquini et al. 2004, and  
Pasquini et al. 2005), one of the possibilities of explaining the presence of Li in
GC stars is that this element, destroyed in the previous  
generations of stars, has been also produced by them, for instance during the
AGB phase, and brought to the surface by the so called Cameron- 
Fowler (Cameron and Fowler 1971) mechanism.

Be cannot be produced in stars, so its mere detection in  
NGC~6752
(as well as in NGC~6397) shows beyond any doubt that the gas which  
formed the
 stars being here studied, has been sitting in a cloud and exposed to the  
GCRs
for a few hundred million years.

In principle, at least for the NGC~6752 stars, one could invoke  
that the chemical
composition we see now is the product of a more recent mixing (or  
mixture) of
'pristine' (Be-rich) and 'processed' (Be-poor) gas (see e.g.  
Charbonnel and Prantzos
2006 for a similar approach).

The observations of NGC~6397 are difficult to be explained simply  
by a dilution
scenario. The large N abundance observed in most stars of this  
cluster indicate
transformations of a fraction of the original O into N, in  
agreement with the only
moderate excess of O (roughly a factor of about 2 smaller than in  
field stars having the
same excess of $\alpha-$elements). This would suggest a similar  
dilution factor for the
"pristine" material in that cluster, requiring a significant Li  
content in the "processed"
material in order to recover the Spite plateau value,  
approximately observed in all
NGC~6397 stars. Such a Li production would in turn favour the AGB  
stars as sources of
the "processed" material. A further requirement is that this  
"processed material" should
also be rich in Be, since the Be content of NGC~6397 is the  
correct one for its [Fe/H]:
this requires exposition to GCRs for a few $10^8$~yrs.  
These coincidences seems
rather hard to accept, mainly if present in several clusters.  

On the other hand, the case of NGC6752 seems easier. In this case,  
the most direct
measure of the dilution of pristine material with processed ones  
in O-poor stars is
provided by a comparison of the abundance of O with that observed  
in the
most O-rich stars in the cluster, that we may assume to be  
composed of "unprocessed"
material. Star 4428 is a good candidate for such group: indeed it  
has an
abundance pattern similar to that of field stars of the same  
metallicity for all
observed elements (Be, Li, CNO, Na, etc., if the N abundance  
provided by our robust
NH observation is adopted). For what concerns stars made of  
"processed" material,
unfortunately,  the O abundance is not available for star 200613;  
however, we may assume for
this star the same dilution factor we would obtain for stars  
having a similar composition,
like e.g. star 4907 (G01, Carretta et al. 2005), 
for which [O/Fe]=--0.25. The difference in  
O content between stars
4428 and 4907 indicates a dilution factor of $\sim 4$. This is  
slightly more than the
difference in the Li content ($\sim 3$); however, this difference  
is likely within
the observational uncertainties (while not excluding some fresh Li  
production). We should
then expect a similar difference in the Be abundances. Our  
spectrum of star 200613 shows only an upper limit, that is fully compatible  
with a similar abundance of
Be. Hence current observations of NGC~6752 may well be explained  
by a variable mixing of
"pristine" and "processed" material, and  the processed  
material could originate both
in AGB stars and massive rotating ones. Again, the presence of Be  
indicates that the
'pristine' material has been exposed to GCRs for a few $10^8 
$~yrs. This strongly
suggest that the material from which this cluster formed followed  
a chemical evolution
similar to those of field stars of similar metallicity.

\subsection{Be abundance variations in the cluster ?}

Due to measurement errors and low S/N of star 200613, our upper limit
value does not allow us to exclude that its Be abundance is
lower than that of star 4428, nor that, instead, the two stars have
a similar Be content. 
We can however firmly exclude that it has
a Be abundance which is a factor of 2
larger.
The determination of a Be abundance difference
between polluted and unpolluted stars
is possibly the only way to understand how many generations of  
stars have
preceeded the one we actually observe.

While it is well established that  GCs have suffered by pollution
from a previous generation of stars in the CNO elements and  
perhaps He
(see e.g. Carretta et al. 2005,  Piotto et al. 2005), it is not  
clear which
progenitor  has been responsible for the cluster chemical  
pollution.  Several authors
(e.g. Ventura et al. 2002, D'Antona et al. 2005) favour  
intermediate mass AGB stars,
while others (Meynet and Maeder 2006, Prantzos and Charbonnel  
2006) consider
massive, rotating  donors.

Both scenarios have in common that the polluters are a second
generation of stars having the same metallicity of the cluster and  
composition of the field, 
which pollute the ISM in light elements but not in Fe.
The ``polluted'' stars were a subsequent (third) generation, born  
from (or contaminated by ) the ejecta of the second generation.

The unpolluted stars should therefore be ``second generation''  
\footnote{ We adopt here the term 
second generation  in a rather loose sense,
simply  to
indicate that GC stars where preceded by several 
generations of stars.}
stars, similar in everything to the
field stars, while the ``polluted'' objects are heavily  
contaminated, third generation.
If this idea, common to both scenarios, were true, then 
we may expect a difference in Be abundance between the two
generations of objects. The third generation of objects
should have a younger age, and therefore a larger
Be abundance. On the other hand the 
polluting material should be Be-depleted 
due to the high temperatures necessary
to form the excess nitrogen.
Therefore there are {\em a priori}
three possibilities:
\begin{enumerate}
\item 
the third generation stars have higher Be abundance;
in this case the age difference is large enough
that a significant amount of Be has built up
and/or the fraction of Be-depleted polluting material
is small enough not to cancel this signature.
\item
the third generation stars have lower Be abundance;
in this case the age difference between 
the second and third generation is small 
and the pollution by Be-depleted material dominates.
\item
the third generation stars have the same Be
abundance as the second generation stars;
this situation may arise in a number
of ways by fine tuning the age difference
between the two generations and the fraction
of Be-depleted polluted material incorporated
into the new stars; ranging from very small 
age difference and little pollution to large
age difference and large pollution.
\end{enumerate}

Our observations clearly rule out
the first of the three cases, but the 
other two remain viable.
The possibility of fine tuning age
and pollution in the third case
is however limited by the observations
of O and Li. From these two elements
the dilution factor is in the range
three to four and should be the same
also for Be.  Therefore, in principle,
if the two stars had the same Be abundance
we could estimate the age difference
using this dilution factor and Fig. 3,
assuming that the total amount of CNO,
necessary for the Be production, has not changed.

If the Be content of star 200613  
were similar to that
of  the unpolluted star 4428, 
for sake of completeness, we should mention that 
there is another scenario which 
could be envisaged:
polluted and unpolluted stars belong to the same generation, and the
metals and the CNO anomalies are produced by the same preceding
generation, but perhaps by stars of different stellar masses. 
The previous generation of
GC stars was extremely metal poor
(perhaps even zero metal),  dominated by  
stars of large masses;
these exploded and ejected mass very  quickly  and
their total  mass was a substantial fraction of the GC total masses.
The Supernovae contributed to the
global enrichment of metals in the Galaxy, without  
leaving any local inhomogeneity,  
while CNO processing and ejection through the  
slower stellar winds, prior to SNe explosions, 
produced the differential enrichment in the cluster of  Na, N, Mg,  
and the depletion of  Li in the cluster gas. 
It should be however noted, that this scenario
requires that the SN explosions do not lead to 
a total mixing of the gas, which would cancel
the inhomogeneities created by the winds.
This may be difficult to reconcile with
the cluster homogeneity in Fe, which is 
better than 10\%. 
Additionally, it might be
difficult to reconcile the required nucleosynthesis with current
understanding of the yields from core collapse SNe.

An accurate determination of the Be abundance in the 
polluted star  200613 
should enable us to discriminate
among competing scenarios.

\balance

\begin{acknowledgements}
{ We are grateful to the referee, R. Cayrel,
for pointing out the consequences of excluding
a large Be abundance in star 200613.}
PB acknowledges support from the MIUR/PRIN 2004025729\_002 and from EU
contract MEXT-CT-2004-014265 (CIFIST).
We
would like to emphasize as these observations are absolutely  
challenging and
we thank the UVES team for building such a
wonderful instrument and the VLT operations for carrying them on.
\end{acknowledgements}
\bibliographystyle{aa}

\end{document}